\documentclass[nonacm]{acmart}
\AtBeginDocument{%
  \providecommand\BibTeX{{%
    \normalfont B\kern-0.5em{\scshape i\kern-0.25em b}\kern-0.8em\TeX}}}

\setcopyright{rightsretained}
\copyrightyear{2024}
\acmYear{2024}
\acmDOI{XXXXXXX}

\acmConference[WWW '24]{WWW 2024}{May 13--17,  2024}{Singapore}
\acmISBN{978-1-4503-XXXX-X/18/06}

\usepackage{balance}
\usepackage{soul}
\usepackage{wrapfig}
\usepackage{hyperref}
\usepackage{graphicx}
\usepackage{subcaption} 
\usepackage{float}
\usepackage{fontawesome}
\usepackage{subcaption}
\usepackage{caption}
\usepackage{tabularx}
\usepackage{colortbl} 

\begin{document}
\title[GET-Tok: A GenAI-Enriched Multimodal TikTok Dataset Documenting the 2022 Attempted Coup in Peru]{GET-Tok: A GenAI-Enriched Multimodal TikTok Dataset  Documenting the 2022 Attempted Coup in Peru}
\author{Gabriela Pinto, Keith Burghardt, Kristina Lerman, Emilio Ferrara}
\affiliation{\institution{University of Southern California} \country{Los Angeles, CA, 90007, USA}}
\email{gpinto@usc.edu, keithab@isi.edu, lerman@isi.edu, emiliofe@usc.edu}






\renewcommand{\shortauthors}{Pinto, et al.}

\begin{abstract}
TikTok is one of the largest and fastest-growing social media sites in the world. TikTok features, however, such as voice transcripts, are often missing and other important features, such as OCR or video descriptions, do not exist.
We introduce the Generative AI Enriched TikTok (GET-Tok) data, a pipeline for collecting TikTok videos and enriched data by augmenting the TikTok Research API with generative AI models.
  As a case study, we collect
  videos about the attempted coup in Peru initiated by its former President, Pedro Castillo, and its accompanying protests. The data includes information on 43,697 videos published from November 20, 2022 to March 1, 2023 (102 days). Generative AI augments the collected data via transcripts of TikTok videos, text descriptions of what is shown in the videos, what text is displayed within the video, and the stances expressed in the video. Overall, this pipeline will contribute to a better understanding of online discussion in a multimodal setting with applications of Generative AI, especially outlining the utility of this pipeline in non-English-language social media. Our code used to produce the pipeline is in a public Github repository: https://github.com/gabbypinto/GET-Tok-Peru.
\end{abstract}

\begin{teaserfigure}
\centering
\begin{subfigure}{0.5\linewidth}
\centering
\captionsetup{font=footnotesize} 
\includegraphics[trim={0 50 0 120},clip,height=90px]{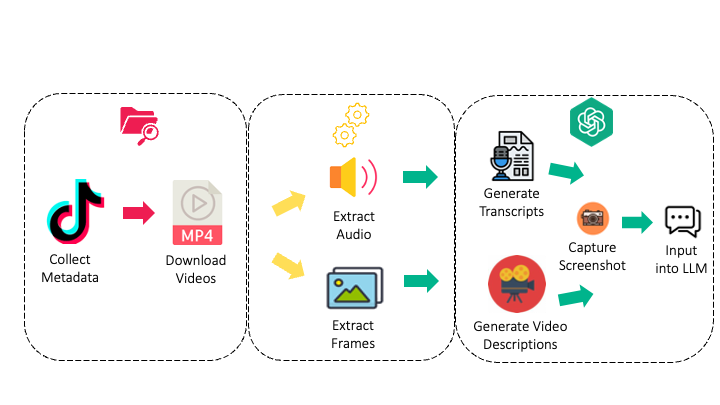}
\caption{GET-Tok: Our framework for Generative AI-Enriched TikTok data.}
\label{fig:sub1}
\end{subfigure}
\begin{subfigure}{0.46\linewidth}
\centering
\captionsetup{font=footnotesize}
\includegraphics[height=90px]{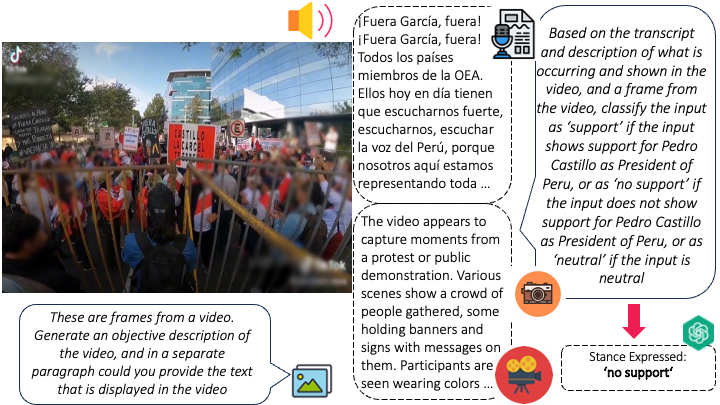}
\caption{Example of classification given transcript, video description and frame.}
\label{fig:sub2}
\end{subfigure}
\caption{(a) Outline of the proposed GET-Tok framework and (b) its application to classify an actual TikTok video}
\label{fig:Outline}
\end{teaserfigure}

\maketitle

\vspace{-10pt} 
\section{Introduction}

Open data initiatives have enabled research on political turmoil and social unrest, predominantly using Twitter streams \cite{davis2016osome}.
Research on TikTok has been understudied, but has included research into U.S. political communication \cite{medina2020dancing,moir2023use}, the spread of COVID-19 information \cite{li2021communicating,shang2021multimodal,basch2021global}, the Brazilian Presidential Election \cite{lima2023use}, disinformation campaigns \cite{alonso2021beyond,espinoza2023propaganda}, or climate change \cite{basch2022climate}.

These studies, however, have highlighted  limitations of TikTok data, questioning if the samples are representative, emphasizing the scarcity of non-English content during political crises in non-English speaking countries, or highlighting limitations in the data or the auto-generated speech transcripts that TikTok APIs produce
\cite{medina2020dancing,purushothaman2022content,li2021communicating,espinoza2023propaganda}. 

These shortcomings motivated us to design the proposed data collection pipeline, shown in Fig.~\ref{fig:Outline}, which includes using generative AI to augment the collected data, providing potential avenues for in-depth analysis of TikTok content on a larger scale. Possible research avenues that could be addressed from the presented work include analyzing content in a multilingual environment, multimodal analysis, potentially enabling answers to social science questions, such as the offline effects of social media, for countries that are traditionally understudied by our research community. We demonstrate a proof-of-concept of this pipeline resulting in the collection of a rich, multimodal dataset of Spanish-based content originating from Peru, documenting a period of social unrest. 

Namely, in November 2022, Peru's Congress attempted to impeach President Pedro Castillo on the charges of leading a criminal organization to profit off government contracts. 
In the face of imminent impeachment, Castillo attempted to dissolve Congress on December 7, 2022, but Congress removed Castillo instead. 
Former Vice President of Peru Dina Boluarte was subsequently sworn in as the new President of Peru. This sparked wide-spread protests, with more than 2,370 demonstrations across the country, including 66 fatalities, according to ACLED \cite{website:ACLED}. The discussion of the political crisis quickly moved to TikTok, which has 20 million users in Peru \cite{Bianchi2024},
providing a case study for the interactions between online activity and offline events.

Using the pipeline described in this paper, we collected 43,697 videos, along with 2,345 Whisper-generated audio transcripts \cite{radford2023robust} and 2,345 GPT-4 generated descriptions of the the content of the videos \cite{achiam2023gpt}. We are sharing the dataset publicly with the research community.\footnote{https://github.com/gabbypinto/GET-Tok-Peru-data}
We are also publishing our data collection pipeline codebase on Github.\footnote{https://github.com/gabbypinto/GET-Tok-Peru}


\section{Data Collection}
\begin{figure}
\centering \footnotesize
\vspace{-.2cm}  
\includegraphics[clip, trim=5 10 20 30, width=\columnwidth]{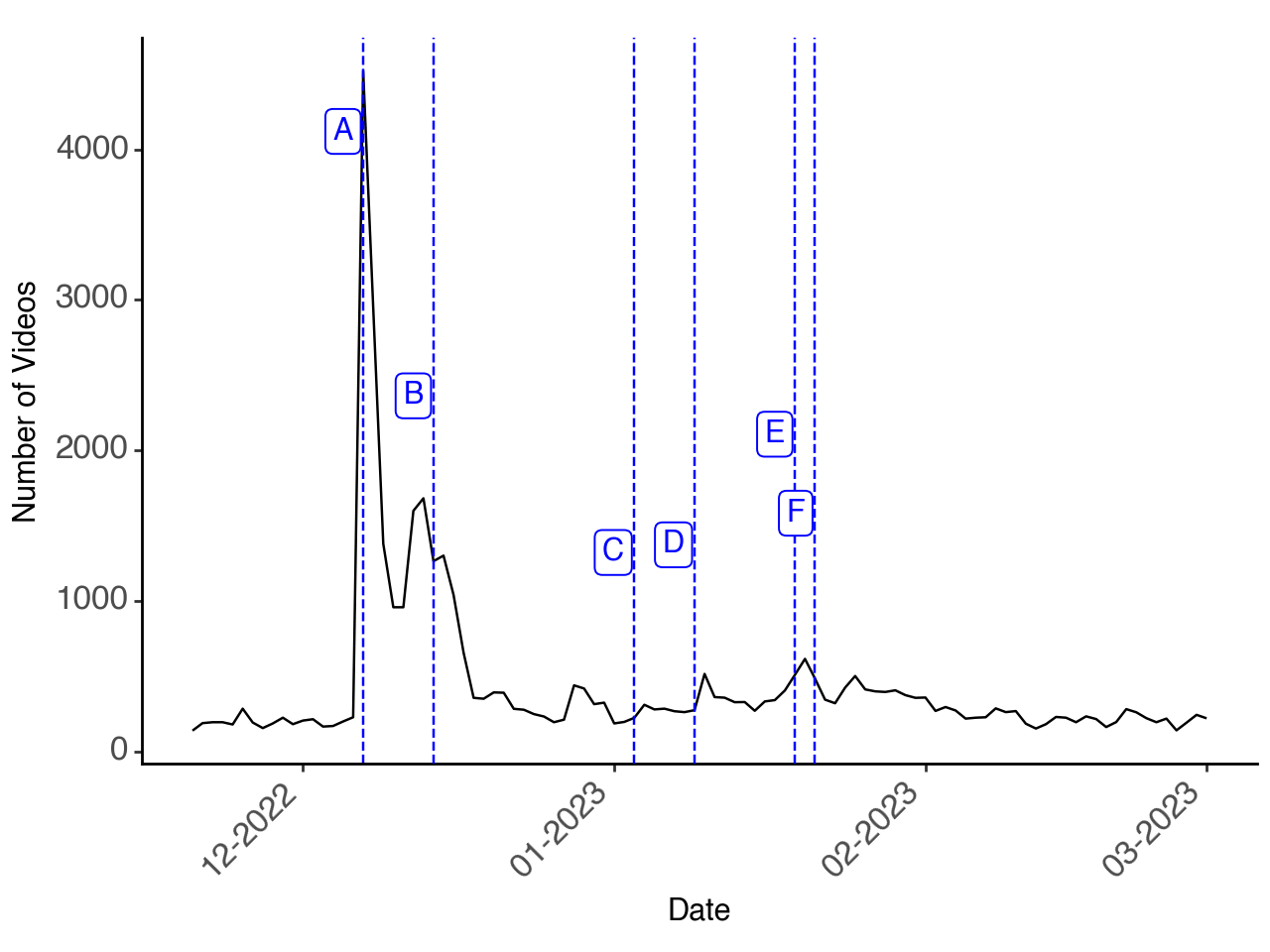}
\begin{tabularx}{\columnwidth}{lll} %
\hline
\textbf{\#} & \textbf{Date} & \textbf{Description} \\ 
\hline

A & 12-07-2022 & President Castillo Attempts to Dissolve Congress \\ 
\hline

B & 12-14-2022 & Boluarte declares State of Emergency \cite{cano2022peru} \\ 
\hline

C & 01-03-2023 & "Great March for Peace" in Cusco \cite{reuters2023conservative} \\ 
\hline

D & 01-09-2023 & 18 Civilians Killed by Police Forces during a Protest in Juliaca \cite{mcdonald2023peru}\\ 
\hline

E & 01-19-2023 & Thousands March in Lima to Demand President's Resignation \cite{schmidt2023peruprotests}\\ 
\hline

F & 01-21-2023 & National Police of Peru (PNP) raid the National University of San Marcos \cite{collyns2023peru}\\ 
\hline
\end{tabularx}
\vspace{-.4cm}
\caption{Timeline of events and volume of TikTok posts.}
\vspace{-.5cm}
\label{fig:postPerDay}
\end{figure}

We continuously query the TikTok Research API\footnote{https://developers.tiktok.com/doc/about-research-api/} based on a set of keywords and hashtags, as shown in Table \ref{tab:keywords_hashtags}, and published between November 20, 2022, and March 1, 2023. The quantity of videos published during the period are shown in Fig.~\ref{fig:postPerDay}. These dates cover videos from 2 weeks before the political crisis until several months after protests reached their zenith.  The collection of videos' features includes their unique identifier, the video's publication time and country code, the author's username, the description of the video, the unique identifier of the music (if any) in the video, the number of \textit{likes, comments, shares, and views} the video received at the time of collection, the list of unique identifiers of the effects (if any) applied in the video, the list of associated hashtags, the unique identifier of its associated playlist (if any), and the audio transcripts (if any). 

After the data collection process, the actual video files are downloaded with the Pyktok Python module\footnote{https://github.com/dfreelon/pyktok}. To download each video, the post's username and ID are extracted and used to create a link with the format \textsl{https://www.tiktok.com/@username/video/id}. 

\begin{table}[t]
\centering
{\footnotesize
\begin{tabular}{|p{4cm}|p{4cm}|}
\hline
\textbf{Keywords} & \textbf{Hashtags} \\
\hline
 Castillo,\ Presidente\ Castillo, Pedro Castillo,\ Dina\ Boluarte,\ Presidente\ Boluarte,\ Boluarte & PedroCastillo,\ Castillo,\ Boluarte,\ pedrocastillo,\ pedrocastilloperu, \ pedrocastilloper\'{u},\ golpedeestado,\ GolpeDeEstado,\ golpedeestadoperu,\ crisispoliticaenperu,\ crisispoliticaenper\'{u} \\
\hline
\end{tabular}
}
\caption{List of Keywords and Hashtags Applied in the Query}
\label{tab:keywords_hashtags}
\vspace{-.5cm}
\end{table}

Between querying the videos from the TikTok API (50,631) and downloading them via Pyktok, 1,427 are privatized and unavailable to download; thus we recommend reducing the time between feature extraction and video download to maximize the content gathered. In addition, Pyktok fails to download posts that contain a slideshow of photos, which represents 1.5\% of the collected videos. Pyktok will also occasionally fail to fully download videos due to internet connectivity issues, which can be resolved by querying the url again after approximately 100 seconds. Futhermore, Pyktok will successfully download but in a format invalid for analysis. As of now, this error is unexplained and appeared in approximately 9.2\% of the data collected. Our dataset is reduced to 43,697 after the removal of the privatized videos, and unsuccessful downloads. Overall, these errors had little impact on our data collection. 

\begin{figure}[t]
\centering
\vspace{-.5cm}
\includegraphics[width=\columnwidth]{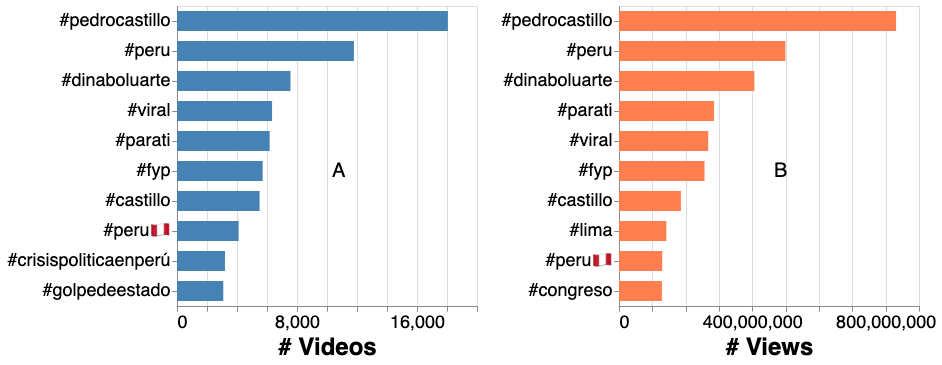}
\caption{Top 10 Most Frequent and Viewed Hashtags}
\label{fig:tophashtags}
\end{figure}

\begin{table}[t]
\centering\footnotesize
\begin{tabular}{|l|c|}
\hline
\textbf{Attribute} &  \textbf{Sum}\\ \hline
No. Users & 17,181  \\ \hline
No. Likes & 63,539,322 \\ \hline
No. Views & 1,396,069,669 \\ \hline
No. Comments & 6,264,238 \\ \hline
No. Shares & 6,964,547 \\ \hline
\end{tabular}
\caption{Statistics on the Dataset}
\label{tab:statistics}
\vspace{-.5cm}
\end{table}

\subsection{Detailed Statistics}
Our dataset has 43,697 videos posted during critical political events during the observed period. Most notably, Pedro Castillo's attempted coup and removal from office were followed by Pro-Castillo protests and then protests against the disproportionate police response. The detailed description of the data collected, all of which have been published within Peru, is shown in Table~\ref{tab:statistics} and Fig.~\ref{fig:tophashtags}.

A small proportion of the videos collected from the collected posts included TikTok-generated transcripts, which has the potential to contain rich-linguistical data. The percentage of videos with its transcripts included, with respect to the publication date, range between 1.4 to 15.3\%. Due to the low proportion of transcripts provided by the TikTok API, we applied OpenAI's Whisper to generate the transcripts for a sample of 2,345 videos. The videos were chosen based on the videos' publication dates. The specific set of dates were chosen to provide a preliminary analysis of the comparison between the two types of transcripts surrounding critical dates within the period. 
\begin{figure}[ht]
\centering 
\includegraphics[width=\columnwidth]{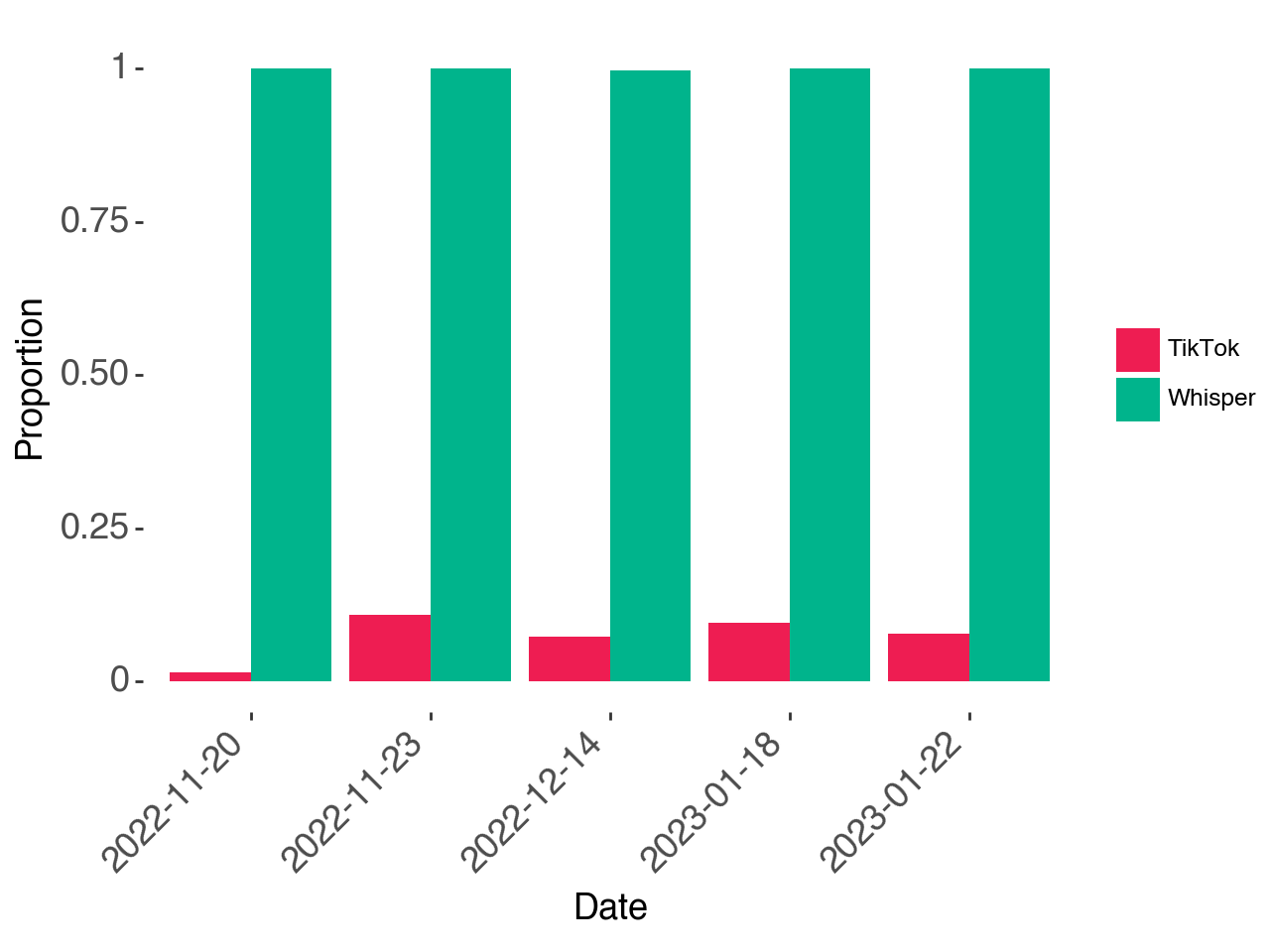}
\vspace{-.5cm}
\caption{Share of video transcripts extracted by TikTok and Whisper.}
\vspace{-.4cm}
\label{fig:transcripts_props}
\end{figure}


\begin{figure}[h]
\centering
{\footnotesize
\begin{tabular}
{|p{3.9cm}|p{3.9cm}|}
\hline
TikTok Transcript & Whisper Transcript \\
\hline
\sethlcolor{yellow}\hl{patriotas} \colorbox{pink}{\phantom{.}} a mi retaguardia está el \sethlcolor{yellow}\hl{palacio} de \sethlcolor{yellow}\hl{gobierno} \colorbox{pink}{\phantom{.}} \sethlcolor{yellow}\hl{la} presidenta actual es la señora \sethlcolor{yellow}\hl{boluarte} \colorbox{pink}{\phantom{.}} \sethlcolor{yellow}\hl{tiene} un gabinete ministerial más técnico\colorbox{pink}{\phantom{.}} \sethlcolor{yellow}\hl{los} gabinetes del señor \sethlcolor{yellow}\hl{castillo} eran más corruptos e ineptos \colorbox{pink}{\phantom{.}} \sethlcolor{yellow}\hl{sin} embargo \colorbox{pink}{\phantom{.}} quedan tentáculos mafiosos en el aparato estatal y deben ser \sethlcolor{cyan}\hl{estirpados} por el actual gobierno \colorbox{pink}{\phantom{.}} \sethlcolor{yellow}\hl{los} patriotas apoyamos a las \sethlcolor{yellow}\hl{fuerzas} \sethlcolor{yellow}\hl{armadas} la \sethlcolor{yellow}\hl{policía} \sethlcolor{yellow}\hl{nacional} del \sethlcolor{yellow}\hl{perú} y la lucha frontal contra la corrupción \colorbox{pink}{\phantom{.}} \colorbox{pink}{\phantom{.}} \sethlcolor{yellow}\hl{viva} \sethlcolor{yellow}\hl{el} \sethlcolor{yellow}\hl{perú} \colorbox{pink}{\phantom{.}} & 

\sethlcolor{yellow}\hl{Patrotas}, a mi retaguardia está el \sethlcolor{yellow}\hl{Palacio} de \sethlcolor{yellow}\hl{Gobierno} \sethlcolor{pink}\hl{.} \sethlcolor{yellow}\hl{La} presidenta actual es la señora \sethlcolor{yellow}\hl{Boluarte} \sethlcolor{pink}\hl{.} \sethlcolor{yellow}\hl{Tiene} un gabinete ministerial más técnico \sethlcolor{pink}\hl{.} \sethlcolor{yellow}\hl{Los} gabinetes del señor \sethlcolor{yellow}\hl{Castillo} eran más corruptos e ineptos\sethlcolor{pink}\hl{.} \sethlcolor{yellow}\hl{Sin} embargo, quedan tentáculos mafiosos en el aparato estatal y deben ser \sethlcolor{cyan}\hl{extirpados} por el actual gobierno\sethlcolor{pink}\hl{.} \sethlcolor{yellow}\hl{Los} patriotas apoyamos a las \sethlcolor{yellow}\hl{Fuerzas} \sethlcolor{yellow}\hl{Armadas}, la \sethlcolor{yellow}\hl{Policía} \sethlcolor{yellow}\hl{Nacional} del \sethlcolor{yellow}\hl{Perú} y la lucha frontal contra la corrupción. ¡Viva Perú!  \\
\hline
\end{tabular}
}
\caption{Transcripts from TikTok and Whisper with yellow for capitalization, red for punctuation, blue for spelling discrepancies.}
\label{fig:transcripts_examples}
\vspace{-.5cm}
\end{figure}

\section{AI-Powered Data Augmentation}
\subsection{Generating Whisper Transcripts}
The transcripts are generated through a two-step process. Initially, the audio is extracted from the video content using FFmpeg, a widely recognized software tool for processing multimedia files.\footnote{https://ffmpeg.org/} Following the extraction of audio, we utilize OpenAI's Whisper \cite{radford2023robust}, to produce the transcripts.\footnote{https://github.com/openai/whisper} Specifically, we opt for the `large-v3' model of Whisper due to its higher accuracy and semantic richness in transcription when compared to the `base' model. To illustrate the effectiveness of our method, we compare transcripts generated by Whisper with those provided by TikTok in Fig.~\ref{fig:transcripts_examples}, highlighting discrepancies in capitalization (yellow), punctuation (red), and spelling (blue). Notably, an instance of misspelling in the TikTok transcript was correctly spelled in the Whisper-generated transcript, showcasing the better performance of the latter. This analysis included posts from specific dates: November 20, 2022; November 23, 2022; December 14, 2022; January 18, 2023; and January 22, 2023. Our limited collection is primarily due to the audio extraction process and Whisper's limited throughput of approximately 19 videos per hour. The comparison of transcript proportions between TikTok and Whisper is depicted in Fig.~\ref{fig:transcripts_props}.

We evaluate the similarity of 182 transcripts through several methods. The transcripts are compared based on their word count, resulting in a Pearson correlation \(r\) of 0.9816 and \(R^2\) of 0.9635. Thus, indicating that the word count between the TikTok and Whisper transcripts were nearly equivalent.  We apply the Jaccard similarity index based on the set of words presented in each transcript, resulting in a mean score of 0.76. To further assess the similarity between the transcripts, we employ BETO embeddings \cite{CaneteCFP2020}---a BERT-based model tailored for Spanish language data. The average cosine similarity was 0.927 indicating a high degree of similarity between the two sets of transcripts. 



\subsection{Generating Video Descriptions with GPT-4}
For the purpose of conducting a comprehensive multimodal analysis, we use GPT-4 to generate descriptions of the activities depicted, as well as any text displayed, within these videos. We specificially use the gpt-4-vision-preview model,\footnote{https://platform.openai.com/docs/models/gpt-4-and-gpt-4-turbo} one of the best models available at the time of writing this paper \cite{achiam2023gpt}.  While OpenAI has not specified the exact dataset used for training their models, it is conceivable that the training data may include information pertinent to the political situation in Peru. This could explain any difference in performance for this pipeline applied to the present dataset compared to any other datasets analyzed in the future.

The description generation sub-pipeline is twofold. We first process the videos using the OpenCV Python package,\footnote{https://pypi.org/project/opencv-python/} a tool that enabled us to capture 1 out of every 300 frames from the videos. These frames are then fed into GPT-4 along with the prompt: \textit{``These are frames from a video. Please generate an objective description of what is happening in the video and, in a separate paragraph, provide any text that appears in the video.''} We were able to generate descriptions for 2,345 videos.

\subsection{GPT-4 Multimodal-Based Stance Detection}
The described pipeline showcases a powerful application in classification tasks, particularly using GPT-4 to assess the stance conveyed by various inputs. This approach is applied to numerous posts, illustrated in Fig.~\ref{fig:sub1}, aiming to categorize each post based on its sentiment towards Pedro Castillo's presidency---whether in support, opposition, or neutral. 

The integration of audio and video frame extraction, alongside generating detailed descriptions of the video content, enables us to achieve significantly more accurate classifications than those based on the typically brief video captions. By employing the comprehensive methodology outlined in Fig.~\ref{fig:sub1}, we can create detailed textual representations of the content. An additional layer of analysis is introduced by including detailed descriptions of the video content generated by GPT-4 as input. 

The classification prompt is as follows: \textit{``Based on the transcript, description of what is occurring and shown in the video, and a frame from the video, classify the input as `support' if the input shows support for Pedro Castillo as President of Peru, or as `no support' if the input does not show support for Pedro Castillo as President of Peru, or as `neutral' if input is neutral.''} 

\begin{figure}[t] 
    \centering
    \begin{minipage}[t]{0.3\linewidth}
        \captionsetup{font=footnotesize}
        \centering\includegraphics[height=4cm,keepaspectratio]{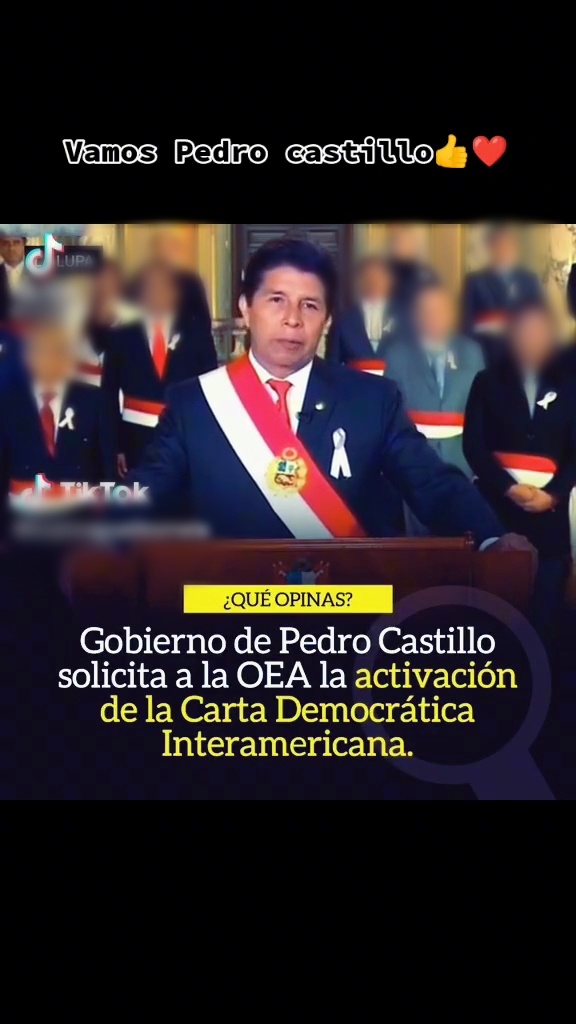}
        \subcaption*{{(A) Vamos\ Pedro\ \\ castillo\ el\ pueblo\ \\ 
        está\ contigo\ \includegraphics[width=1em]{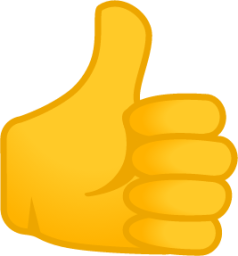}\ \#parati\ \#viralvideo\
        \#tiktok\ \#peruanadas\ \#love\
        \#pedrocastillopresidente2021 \textit{Translation: Let's go Pedro Castillo the people are with you} \textit{Translation (screen): Let's go Pedro Castillo}}}
    \end{minipage}
    \hfill
    \begin{minipage}[t]{0.3\linewidth}
        \captionsetup{font=footnotesize}
        \centering
        \includegraphics[height=4cm,keepaspectratio]{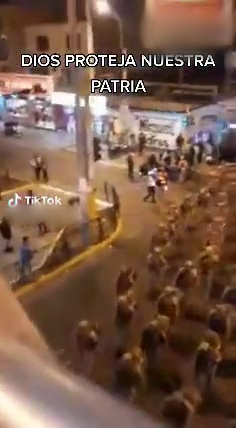}
        \subcaption*{(B) \#antaurohumala \#huanta \#pedrocastillo \#cierredelcongreso \#golpedeestado \#insurgencia \#andahuaylas \#peru \#peruanos\\
        \textit{Translation (screen): God protect our homeland}}
    \end{minipage}
    \hfill
    \begin{minipage}[t]{0.3\linewidth}
        \captionsetup{font=footnotesize}
        \centering
        \includegraphics[height=4cm,keepaspectratio]{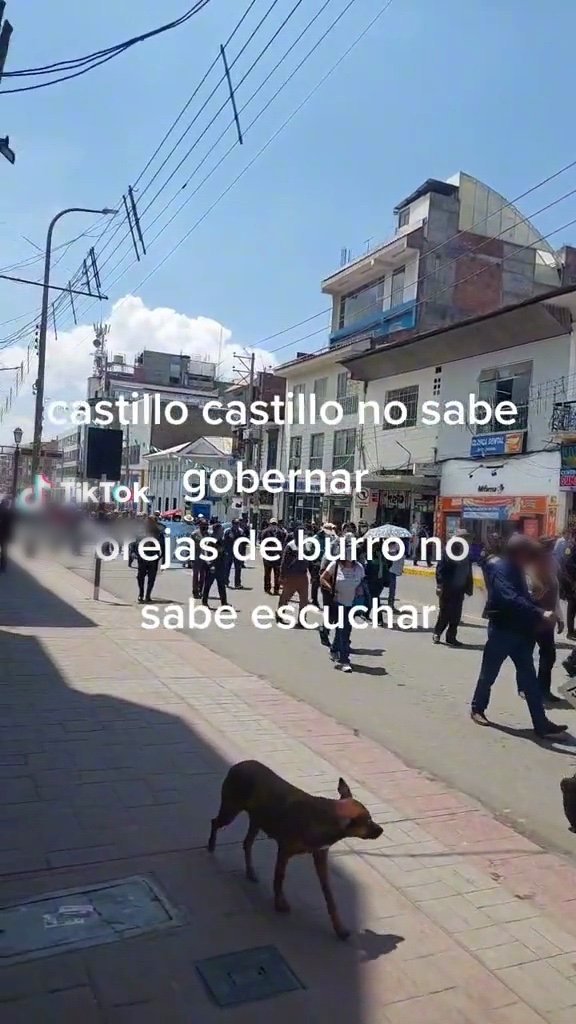}
        \subcaption*{(C) castillo orejas de burro\\
        \textit{Translation: Castillo donkey ears.}\\
        \textit{Translation (screen): Castillo Castillo doesn't know how to govern\\ Donkey ears doesn't know how to listen.}}
    \end{minipage}
    \hfill
    \vspace{-.35cm}
    \caption{Videos and their captions separated by the detected stance: (A) `Pro-Castillo' (B) `Neutral' (C) `Anti-Castillo'}
    \label{fig:screenshots}
    \vspace{-.5cm}
\end{figure}

A few examples of the inputs, along with their corresponding ``video\_description,'' provided by the API, are shown in Fig. ~\ref{fig:screenshots}. The examples suggest good agreement with expectations; in future work,  we will use human annotation to quantify this agreement.

\section{Conclusions}
In summary, this study presents a novel approach to collecting and analyzing social media content, focusing on TikTok videos from the political crisis in Peru. Utilizing generative AI, including the Whisper and GPT-4 models, we crafted a comprehensive pipeline that enhances the collected data with detailed audio, video, and textual analysis. The result is a valuable multimodal dataset of 43,697 videos, equipped with enhanced transcripts and video descriptions. This dataset bridges a significant research gap by focusing on non-English content and regions typically overlooked in social media studies. Our methodology showcases the potential of AI in extracting nuanced insights from social media narratives, contributing to a deeper understanding of the relationship between online discourse and real-world events. By sharing our dataset and codebase, we encourage further exploration into the impacts of digital platforms on societal and political dynamics, highlighting the critical role of AI in advancing social science research.

\paragraph{Limitations}

The dataset sheds light on Peru's political crisis discussions online but has limitations. It doesn't fully capture the sentiment on former President Castillo's attempted coup, considering Peru's 68.68\% social media penetration and approximately 20 million TikTok users out of a 34.5 million population. This study's methodology introduces GPT-4 models for data augmentation, which, despite not being open-source or free, offers a new approach. However, Whisper encounters issues with background noise, omits emojis from the source videos and AI-generated video descriptions present ethical dilemmas due to potential biases.

\paragraph{Ethical Considerations}
In compliance with the TikTok Research API Terms of Service, we regularly update and prune our dataset, ensuring its alignment with current API offerings and adherence to TikTok's guidelines. Shifting focus, our study employs GPT-4 and Whisper to generate textual and visual content analyses. OpenAI's nondisclosure of the models' training data raises ethical concerns about potential biases. Acknowledging this, we aim to prioritize bias detection and mitigation in our future work, enhancing the research's ethical foundation.

\smallskip \small \textbf{Acknoledgements}. This work was supported in part by DARPA (contract no. HR001121C0169).

\balance
\bibliographystyle{abbrvnat}
\bibliography{references}


\end{document}